\documentclass[%
 reprint,
groupedaddress,
unsortedaddress,
nofootinbib,
 amsmath,amssymb,
 aps,
floatfix,
]{revtex4-2}

\usepackage{graphicx}
\usepackage{dcolumn}
\usepackage{bm}
\usepackage{url}
\usepackage{enumitem}
\usepackage{amssymb}
\usepackage{hyperref}
\usepackage{multirow}
\usepackage{float}
\usepackage{natbib}
\hypersetup{
    colorlinks=true,
    linkcolor=blue,
    filecolor=magenta,      
    urlcolor=blue,
    pdftitle={Overleaf Example},
    pdfpagemode=FullScreen,
    }
\usepackage[no-test-for-array]{nicematrix}

\begin{document}

\preprint{APS/123-QED}

\title{Are CMB derived cosmological parameters affected by foregrounds associated to nearby galaxies?}

\author{Facundo Toscano}
 \email{facundo.toscano@mi.unc.edu.ar}
\affiliation{%
 Instituto de Astronomía Teórica y Experimental (IATE), CONICET-UNC, Córdoba, Argentina
}%
\author{Frode K. Hansen}%
 \email{f.k.hansen@astro.uio.no}
\affiliation{%
 Institute of Theoretical Astrophysics, University of Oslo, PO Box 1029 Blindern, 0315 Oslo, Norway\\}
 \affiliation{Instituto de Astronomía Teórica y Experimental (IATE), CONICET-UNC, Córdoba, Argentina}
 \author{Diego Garcia Lambas}
  \email{diego.garcia.lambas@unc.edu.ar}
 \affiliation{%
 Instituto de Astronomía Teórica y Experimental (IATE), CONICET-UNC, Córdoba, Argentina\\
 Observatorio Astronómico de Córdoba (OAC), UNC, Córdoba, Argentina\\
 Comisión Nacional de Actividades Espaciales (CONAE), Córdoba, Argentina
}%
 \author{Heliana Luparello}
\affiliation{%
 Instituto de Astronomía Teórica y Experimental (IATE), CONICET-UNC, Córdoba, Argentina
}%
 \author{Pablo Fosalba}
\affiliation{Institut d’Estudis Espacials de Catalunya (IEEC), Edifici RDIT, Campus UPC, 08860 Castelldefels, Barcelona, Spain \\ 
Institut de Ciencies de l’Espai (IEEC-CSIC), Campus UAB, Carrer de Can Magrans, s/n Cerdanyola del Vallés, 08193 Barcelona, Spain}
 \author{Enrique Gaztañaga}
\affiliation{Institute of Cosmology \& Gravitation, University of Portsmouth, Dennis Sciama Building, Burnaby Road, Portsmouth PO1 3FX, UK \\
Institut de Ciencies de l’Espai (IEEC-CSIC), Campus UAB, Carrer de Can Magrans, s/n Cerdanyola del Vallés, 08193 Barcelona, Spain \\
Institut d’Estudis Espacials de Catalunya (IEEC), Edifici RDIT, Campus UPC, 08860 Castelldefels, Barcelona, Spain}

\date{\today}

\begin{abstract}
  We perform cosmological parameters estimation on Planck Cosmic Microwave Background (CMB) maps masking the recently discovered foreground related to nearby spiral galaxies. In addition, we also analyse the association between these foreground regions and recent claims of cosmological causal horizons in localized CMB parameter estimates. Our analysis shows consistent cosmological parameter values regardless of the masking approach, though reduced sky areas introduce larger uncertainties. By modelling the new extragalactic foreground, we identify a resemblance with local parameter variation maps with a statistical significance at the $3\sigma$ level, suggesting that a simplified foreground model  partially accounts, $(40-50) \%$ correlation with $15\%$ uncertainty, for the observed causal horizons. These findings add new evidence to the existence of the new foreground associated with large spiral galaxies and show that estimates of cosmological parameters on smaller patches on the sky can be largely affected by these foregrounds, but that the parameters taken over the full sky are unaltered.
\end{abstract}

\maketitle


\section{\label{sec:intro} Introduction}

The existence of tensions between different estimates of the Hubble constant $H_0$ and the mean mass density fluctuation $\sigma_8$ has been at a long-standing state of debate \cite[for recent reviews]{DiValentino2021, Tully2023, Verde2023}. These tensions could rely on astrophysical biases in the Late Universe associated to systematic errors in standard candles, such as Cepheids and Supernovae \cite{Freedman2024, Riess2024, Scolnic2024}. However, other foreground contamination sources in the Cosmic Microwave Background (CMB) can not be completely excluded. A more radical approach based on new physics has also been discussed in several works \cite{Hu2023, Kamionkowski2023}.

A recent series of papers have shown the existence of a new extragalactic foreground associated mainly to large spiral galaxies \cite[hereafter L2023]{Luparello2023}\cite{Cruz2024}. This extragalactic foreground is detected at the $5.7 \sigma$ level \cite[hereafter H2024]{Hansen2024} (see also \cite{Addison2024} for a discussion of the significance) and could explain several CMB anomalies \cite[hereafter H2023]{hansen2023}\cite[hereafter GL2023]{Lambas:2023gzy}. L2023 and H2024 show that there is a significant correlation between areas of low CMB temperature and the position of large spiral galaxies. This effect extends well beyond the galaxy halo up to $\sim 4$Mpc/h in projection. The large angular extension of this effect shows that the phenomenon is associated to large--scale structures with bright spirals behaving as suitable tracers.

It is of great interest to understand if this new foreground could be affecting the determination of cosmological parameters from CMB data. Here we will re-estimate these parameters using maps from the latest Planck Public Release 4 \cite{PlanckPr4}, taking into account the new foreground. We first construct a specific analysis pipeline that is validated with the official Planck results. We then study the impact of the foreground regions by considering three different masks which restrict the analysis to foreground free areas of the CMB.

We will also investigate the role of the new foreground related to the existence of causal horizons defining regions of the sky with largely anomalous cosmological parameter values. These horizons in the observed CMB \cite{Fosalba} (hereafter, FG2021) cannot be accounted for in $\Lambda$CDM synthetic maps, being a rare event with a high statistical significance. Possible explanations have been discussed in \cite{Gaztañaga1,Gaztañaga2,Gaztañaga3,Gaztañaga4,Gaztañaga5}. In H2023, it was found that the angular distribution of nearby galaxies coincide to a large degree with the position of these observed horizons. We study the possible correlation between three different foreground models and the maps of local cosmological parameter estimates in $H_0$ and $\Omega_ch^2$ (from FG2021). For this purpose, we first calculate the correlation using the actual CMB and galaxy data and then estimate uncertainties through both CMB local parameter simulations and foreground models based on mocks galaxy catalogues. Following L2023, H2023 and H2024, we adopt the foreground regions using the 2MASS Redshift Survey (2MRS) \cite{2mrs}. We base the mock foreground models on the MultiDark Planck 2 simulation (MDPL2) \cite{Multidark}.

The paper is organised as follows: Section \ref{sec:metho} describes our analysis pipeline, the analysis choices taken and its validation with official data. Section \ref{sec:Results} presents our main results including the foreground masking analysis and the relation between the foreground regions and the cosmological horizons. Finally, Section \ref{sec:Conclusions} offers a brief summary of our main results and their implications.

\section{\label{sec:metho} Methodology and Data Analysis}

\subsection{Parameter estimation}

In order to explore for possible variations of the cosmological parameters due to the recently discovered extragalactic foreground, we need to select CMB maps and apply appropriate masks for the analysis. We use the Planck Public Release 4 (Planck PR4)\footnote{Data available in \url{https://pla.esac.esa.int/}}, which is the latest reprocessing of both LFI and HFI instruments data using a common pipeline, NPIPE \cite{PlanckPr4}. Due to the joint analysis, there are lower levels of noise and systematic in both frequency and component maps at essentially all angular scales. To mask foreground residuals around the Galactic plane as well as bright extragalactic point sources, we apply the Planck common mask based on the union of the uncertainties of four different foreground subtraction methods used in the Planck Public Release 3 (Planck PR3) \cite{PlanckPr3I} analysis \cite{compsep2018} and leaves about $78 \%$ of the sky area suitable for statistical studies.

The preferred method to estimate the cosmological parameters avoiding these new foreground regions would be their characterisation through a physically motivated model. As there is not yet such a quantitative model that allows us to characterise them accurately (H2023 for more details), we choose to mask projected areas in the CMB map around the large spiral galaxies associated with the foreground and study the parameter changes in the remaining regions. We create different galaxy samples depending on main characteristics such as redshift, size or morphological classification and use these as a basis for a set of different foreground masks. We then in turn analyse the regions outside these different masks.

For creating the masks one needs to find a trade-off between masking as much as possible of the foreground signal, but at the same time leave as much sky fraction as possible to reduce error bars on the cosmological parameters. One important aspect to consider is the size of the masking patch around each galaxy. Since dark matter may play a role in this foreground, we must extend the mask beyond the observed angular size of the galaxy considering the surrounding dark matter halo. This is supported by the observed extent of the temperature profiles (see L2023 and H2023). We know that the foreground signal is stronger in galaxies in dense environments (L2023) and in particularly in galaxy groups (H2024). We also know that the larger spirals contribute more than the smaller spirals. By masking mainly galaxies in groups or larger galaxies, one may reduce a significant part of the foreground signal, but at the same time keep a significant sky fraction for analysis. Furthermore, the more distant galaxies have much smaller angular extensions, allowing to mask a larger physical area around the galaxy, but at the same time one need to take into account that the distant galaxies are more numerous than the nearby galaxies. Although the foreground has not yet been detected independently for $z>0.02$, we expect the signal to be present also for these more distant galaxies and we apply masks out to $z<0.035$.

\begin{figure} 
    \centering
    \includegraphics[width=\columnwidth]{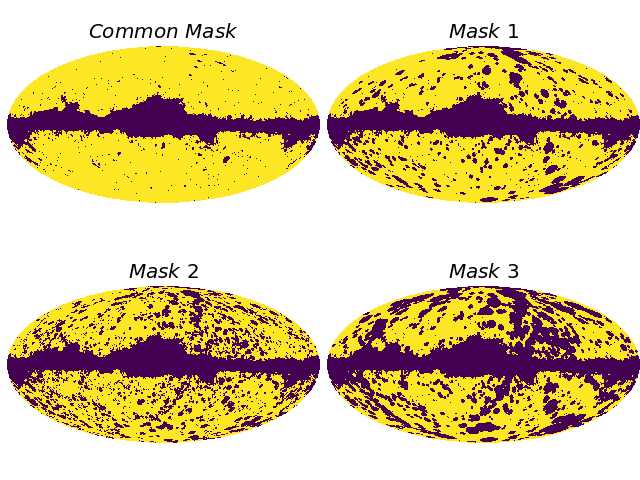}
    \caption{Planck PR3 common mask together with the foreground masks used for the estimation of the cosmological parameters. Each mask differs in size, type and environment of the foreground galaxies considered as well as size of the masking patch (Table \ref{mask_table} for more details).}
    \label{fig:all_masks}
\end{figure}

The three masks which we have chosen are shown in Table \ref{mask_table} and reflect how we have tried to balance all these different concerns. Taking into account that the Planck PR3 common mask has an available sky fraction of $78 \%$ (Figure \ref{fig:pr4_map}), our extended masks increase the covered areas, resulting on $67 \%$, $60 \%$ and $51 \%$ of available sky fraction, respectively. This can be seen in Figure \ref{fig:all_masks}.

\begingroup
\setlength{\tabcolsep}{8pt} 
\renewcommand{\arraystretch}{1.5} 
\begin{table*}
  \centering
   \begin{tabular}{|c|c|c|c|c|}
    \hline
    Mask & Redshift Range & Galaxy Size (GS) & Environment/Type & Masking Patch Size \\
    \hline
    \multirow{2}{*}{Mask 1} & $[0.004, 0.02]$ & GS $> 15$ kpc & Late Spirals in groups & $2$ Mpc \\
    & $[0.02, 0.035]$ & GS $>20$ kpc & Late Spirals in groups & $4$ Mpc \\
    \hline
    \multirow{2}{*}{Mask 2} & $[0.004, 0.02]$ & GS $> 8.5$ kpc & Late Spirals & $1$ Mpc \\
    & $[0.02, 0.035]$ & GS $>20$ kpc & Late Spirals in groups & $4$ Mpc \\
    \hline
    \multirow{3}{*}{Mask 3} & $[0, 0.004]$ & GS $> 8.5$ kpc & Late Spirals & $0.7$ Mpc \\
    & $[0.004, 0.02]$ & GS $> 8.5$ kpc & Late Spirals in groups & $2$ Mpc  \\
    & $[0.02, 0.035]$ & GS $>20$ kpc & Late Spirals in groups & $4$ Mpc  \\
    \hline
   \end{tabular}
  \caption{Characteristics of the masks used for the analysis of cosmological parameters.}
  \label{mask_table}
\end{table*}
\endgroup

\subsubsection{\label{subsec:pipeline} Analysis Pipeline}

\begin{figure} 
    \centering
    \includegraphics[width=\columnwidth]{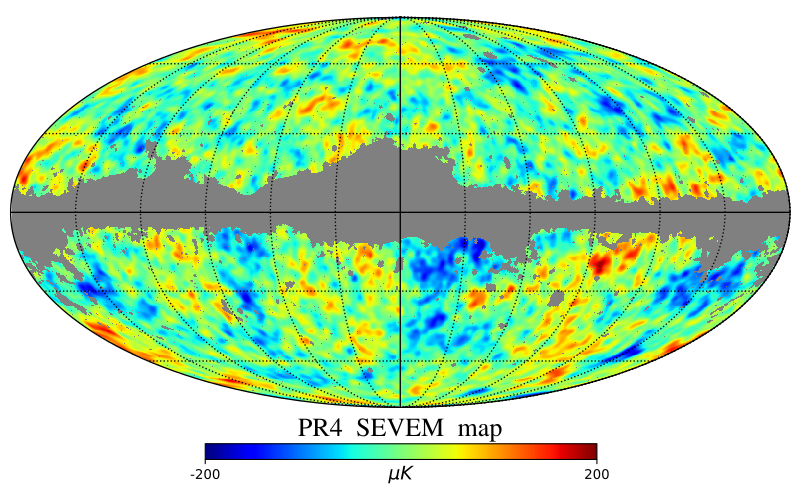}
    \caption{Planck PR4 SEVEM CMB map in galactic coordinates. It is obtained through the combination of both detectors $A$ and $B$. The PR3 common mask with $\sim 78\%$ of available sky fraction is also shown. The map is smoothed with a FWHM of $1^{\circ}$ for better visualisation.}
    \label{fig:pr4_map}
\end{figure}

Our analysis pipeline can be summarized as follows:

\begin{enumerate}[itemsep=4pt, parsep=0.8pt]

\item We first measure the pseudo angular power spectrum $pseudo-C_{\ell}s$ of each masked map. In order to avoid a possible noise bias, we use the cross-spectrum between detector sets A and B of the PR4 maps cleaned with the SEVEM component separation method (see \cite{compsep2018} for details). In Figure \ref{fig:pr4_map} the map obtained through the combination of both detector maps ($0.5 \cdot A+0.5 \cdot B$) together with the Planck PR3 common mask can be seen.

\item Following the approach used for the Planck high-$\ell$ likelihood \cite{PlanckPr3V} and in \cite{Fosalba}, we limit the multipole range to $\ell = [32,2012]$ to minimise galactic foreground residuals at low multipoles and point source contamination as well as instrumental noise at high multipoles. Also, this choice allows the use of a Gaussian assumption for our likelihood and a one-parameter parametrisation ($A_{PS}$) of the point source residuals. To estimate the full-sky angular power spectrum from the masked  $pseudo-C_{\ell}s$ we apply the \texttt{MASTER} approach \cite{master} to obtain band power estimates $\hat C_b$ in bins $b$ of $\Delta\ell=30$ multipoles. As discussed in \cite{Fosalba} this bin-size allows the diagonal approximation of the covariance matrix.

\item In order to estimate the errors on the band power estimates $C_b$, we repeat the same procedure on the 600 publicly available Planck PR4 CMB simulations for each masked map. In PR4, only the SEVEM component separation method has such a large number of corresponding simulations available.

\item Finally, we calculate the best fit $\Lambda$CDM cosmological parameters for each mask by maximizing the likelihood using the multi-parameter minimizer code \texttt{IMINUIT} \footnote{\url{https://scikit-hep.org/iminuit/}}. This approach has been widely used in cosmological parameters estimations using CMB data \cite{Planck2013XVI, CAMEL, Fosalba} and, compared to the well-known Markov Chain Monte Carlo (MCMC) methods, \texttt{IMINUIT} has the advantage of being orders of magnitude faster than traditional Bayesian methods. This gain in speed allows us to optimise the analysis on the various masks used (together with their simulations), analysis that would take much longer with MCMC methods.
\end{enumerate}

In this work, we focus on the basic flat-space $\Lambda$CDM cosmological parameters ($\Omega_c h^2$, $\Omega_b h^2$, $H_0$, $n_s$, $A_s$) and the derived parameter $\sigma_8$, including a foreground residual parameter $A_{PS}$ which consider the combined contribution from the Cosmic Infrared Background (CIB) and unresolved extragalactic point sources \cite{PlanckPr3V}. If one set the limit to $\ell < 2000$, these residuals behave as a single 'Shot-noise' contribution at the power spectrum level and can be well modelled \cite{Planck2015XI}. We fix all the other parameters: $\tau = 0.0522$, $\Sigma m_{\nu} = 0.06$ eV, $N_{eff}=3.04$ and $r=0$. As we are analysing only temperature maps, fixing the optical depth does not cause problems because it is correlated to the primordial amplitude parameter $A_s$ which is free in our analysis. Note that fixing $\tau$ results in smaller error bars for $A_s$ than in the official Planck release were both parameters are estimated jointly.

\subsubsection{\label{subsec:validation} Pipeline Validation}
We validate our pipeline by comparing the official Planck PR3 results for the flat-space $\Lambda$CDM cosmological parameters and our results obtained through the likelihood estimation using \texttt{MASTER} for the spectrum calculation and \texttt{IMINUIT} for the parameters estimation. We restrict our analysis to temperature data and low multipole polarisation for the Planck Collaboration results (TT + LowE in the Planck Release notation).

\begin{figure} 
    \centering
    \includegraphics[width=\columnwidth]{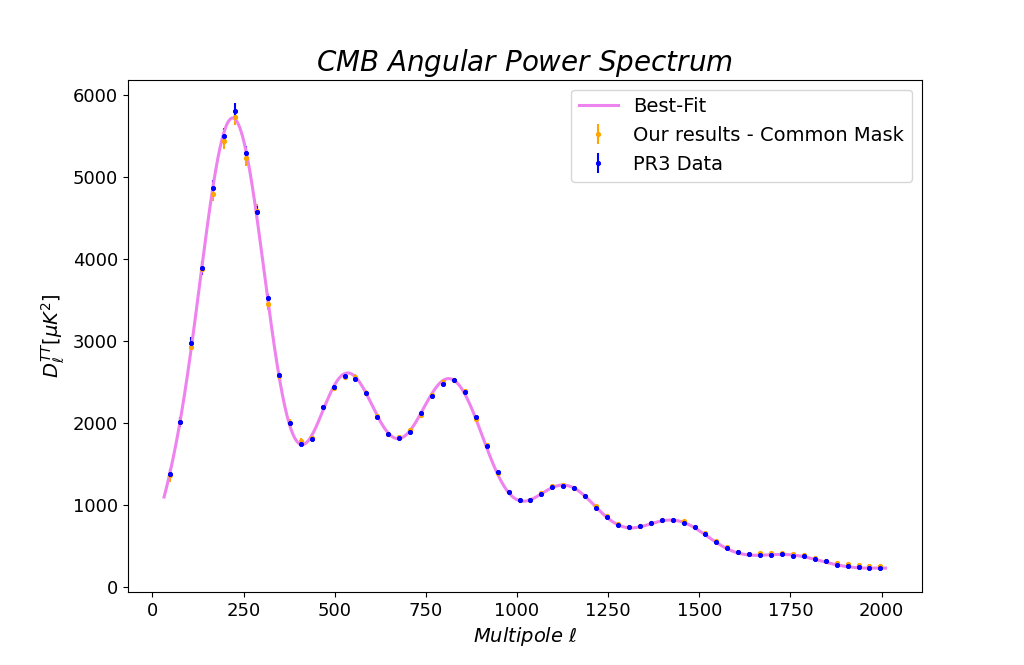}
    \caption{Angular power spectrum for Planck PR3 data (blue), our estimate (orange) and Planck best fit model using \texttt{CAMB}\footnote{\url{https://camb.readthedocs.io/en/latest/}} (pink).}  
    \label{fig:spectrums}
\end{figure}

In Figure \ref{fig:spectrums} we show the angular power spectrum for Planck PR3 data, in blue, and our results, in orange. Also, we show the best fit parameter model for the angular power spectrum in pink. In Table \ref{tab:pipe_valid} we show the comparison of the cosmological parameters results between the official Planck PR3 data (second column) and our analysis (third column). It can be seen that using our simplified pipeline, we obtain a suitable agreement within $1\sigma$ with respect to the Planck temperature data and low multipole polarisation results (see the first column of Table 2 of \cite{PlanckPR3VI} for details).

\begingroup
\setlength{\tabcolsep}{5pt} 
\renewcommand{\arraystretch}{1.4} 
\begin{table} [H]
    \centering
    \begin{tabular}{|c|cc|}
    \hline
        Parameter & PR3 Results  & Our Results \\
        \hline
        $\Omega_c h^2$ & $0.121 \pm 0.002$ & $0.120 \pm 0.002$ \\
        $\Omega_b h^2$ & $0.0221 \pm 0.0002$ & $0.0220 \pm 0.0002$ \\
        $H_0$ & $66.9 \pm 0.9$ & $67.1 \pm 0.8$  \\
         $n_s$ & $0.963 \pm 0.006$ & $0.962 \pm 0.006$   \\
        $10^9 A_s$ & $2.09 \pm 0.03$ & $2.09 \pm 0.01$ \\
        $A_{PS}$ & -- & $57 \pm 4$ \\
        $\sigma_8$ & $0.812 \pm 0.009$ & $0.809 \pm 0.008$ \\
        \hline
    \end{tabular}
    \caption{Comparison between official Planck PR3 results for TT + LowE and our pipeline using only the Planck common mask and 100 simulations for uncertainties.}
    \label{tab:pipe_valid}
\end{table}
\endgroup

\subsection{\label{subsec:horizons} Foreground galaxies and their relation to causal horizons}

In the second part of this paper, we investigate the relation between the extragalactic foreground map models and the causal horizons, defined in FG2021 as regions in the CMB with anomalous cosmological parameters values. In FG2021, the authors investigate CMB temperature maps in an effort to discover regions where cosmological parameters have anomalous values. In aiming to determine the local variations of cosmological parameter estimates across the sky, they construct discs of $60^{\circ}$ centred in low-resolution \texttt{HEALPix} pixels ($N_{side} = 4$, i.e, 192 pixels across the sky). However, for the size estimation of these horizons in the parameter maps they use finer resolution, particularly $N_{side}=32$ (12888 discs across the sky). They use the multipole range $\ell =[32, 2000]$ with a binning of $\Delta \ell =30$ following the official Planck analysis papers. For the error estimation, they re-scale those from Planck power spectrum with their effective area used, i.e $\Delta C_l^D / \Delta C^{Planck} = \sqrt{f_{sky}^{Planck}/f_{sky}^D}$, where $f_{sky}^{Planck} = 0.57$ and $f_{sky}^D \approx 0.05$ corresponds to the sky fraction of the $60^\circ$ discs.

Finally, for the cosmological parameters estimation they use \texttt{IMINUIT} to maximize the likelihood. With this method, they found three regions (horizons) across the sky with anomalous values of the cosmological parameters (see their Figure 28 for more details). In particular, Horizons 1 and 3 have higher values of $H_0$ and $\Omega_ch^2$ than reported in Planck PR3. The parameters $H_0$ and $\Omega_ch^2$ are the parameters most significantly deviating from the full-sky results, coinciding with values obtained by low redshift methods and in tension with the CMB results.

\begin{figure} 
    \centering
    \includegraphics[width=\columnwidth]{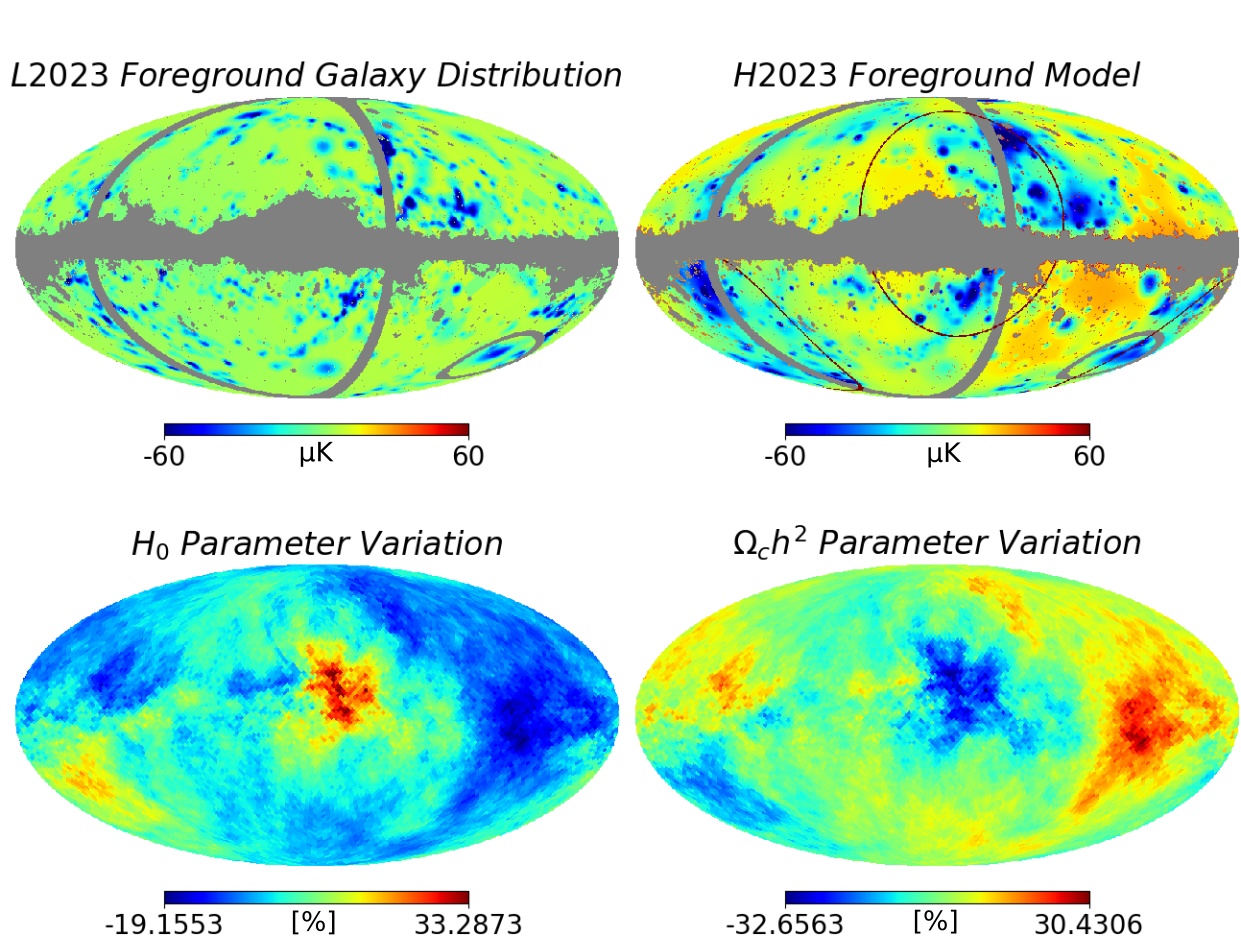}
    \caption{In clockwise sense, the top panels show the large spiral galaxy distribution from L2023 and the galactic foreground model from H2023, where the red circles in the model map correspond to the Horizons 1 and 3 of FG2021. The lower panels show the $\Omega_ch^2$ and $H_0$ parameter maps studied in FG2021 where the scale corresponds to the variation with respect to Planck PR3 values.}
    \label{fig:relation_h13_forg}
\end{figure}

On the other hand, as mentioned in H2023, the regions in the sky where these horizons are present appear to a large degree to overlap with the foreground regions of high galactic density, suggesting that the cosmological parameter values inside the horizons might be affected by the extragalactic foreground of nearby galaxies (Figure \ref{fig:relation_h13_forg}). Taking this into account, we analyse three simple phenomenological models to study and determine the existence and the significance of a possible correlation between this extragalactic foreground and the casual horizons.

\subsubsection{Foreground Models} \label{foreground_models}

\begin{figure*}
    \centering
    \includegraphics[width=0.85\textwidth]{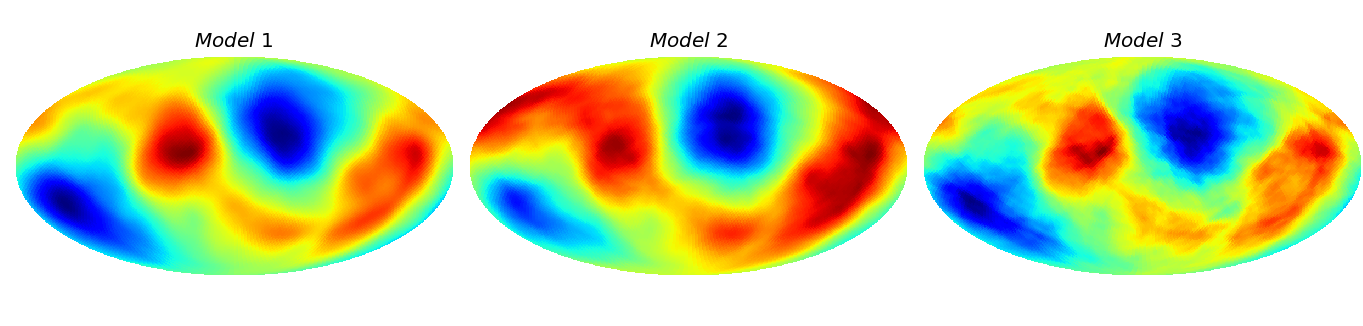}
    \caption{Extragalactic foreground models used for the correlation test with the FG2021 cosmological parameter variation maps. The models have been smoothed with a $60^\circ$ top-hat beam to be suitable for correlation tests with local cosmological parameter variation maps where parameters have been estimated over $60^\circ$ discs. Blue areas represent high galactic density and thereby low model temperature.}
    \label{fig:models_param}
\end{figure*}

Given the lack of a physically motivated foreground model, we have adopted a phenomenological approach considering different temperature profiles around late type galaxies, consistent with the mean values of the observations. We have taken into account the large late type galaxies in three redshift ranges up to $z=0.035$ with a distance threshold ($d_5$). $d_5$ corresponds to the minimum distance for each galaxy to its fifth neighbour within $\Delta z = 600 $km/s. We impose to the galaxies within a $d_5 < d_{5MAX}$ a linear temperature decrement profile with an amplitude at the centre of the galaxy and a profile extent given in Table \ref{tab:models_table}. The corresponding sky maps of these models are shown in Figure \ref{fig:models_param}.
We also applied a $60^{\circ}$ top-hat smoothing in order to have full sky maps suitable to correlate with the cosmological parameters maps of FG2021 which are based on parameter estimates on $60^{\circ}$ discs.

The selected properties of the models are based on the observationsof L2023 and H2023 with temperature profiles around galaxies in agreement with the mean values observed in Planck data. We notice that these mean values can be reproduced, either by localized and deep, or shallow and extended model profiles around each galaxy. While the former models seem more realistic in terms of possible physical processes of interacting materials around galaxies, the latter generally gives a better fit to the data. The real signal may be something in between: if the interacting material is not only bound to the galaxy but also distributed around in galactic filaments extending several Mpc around the galaxies, one could possibly model the signal as a combination of the two approaches.

\begingroup
\setlength{\tabcolsep}{4pt} 
\renewcommand{\arraystretch}{1.2} 
\begin{table*}
    \centering
    \begin{tabular}{|c|c|c|c|c|}
    \hline
        Model & Redshift Range & $d_5$ & Amplitude & Extent \\
        \hline
         Model 1 & $[0.004, 0.02]$ & 3 Mpc & 2$\mu$K & 10 Mpc \\
         \hline
         Model 2 & $[0.004, 0.035]$ & $[0,3.5]$ degrees & [0, 160] $\mu$K &  $[0, 50]$ degrees \\
         \hline
         Model 3 & $[0.004, 0.02]$ & 3 Mpc & 15$\mu$K & 1 Mpc \\
         \hline
    \end{tabular}
    \caption{Characteristics of the foreground models. The parameter $d_5$ is the maximum distance to the 5th galaxy in order for the galaxy to be assigned a foreground signal. Galaxies with higher $d_5$ are related with less dense environments and are not assigned a temperature decrement in agreement with observed properties of the foreground. Amplitude refers to the central temperature decrement. Note that Model 2 corresponds to the foreground model in H2023 extended to higher redshifts. As described in detail there, in this model the galaxy profile amplitude has a quadratic dependence on galaxy size and a power law index $3.5$ dependence on $d_5$. Only galaxies at high density environments, $d_5 < 3.5^{\circ}$, where given a profile. The amplitude was normalized to $-30 \mu$K for galaxies with size $8.5$ kpc and $d_5 = 3.5^{\circ}$.}
    \label{tab:models_table}
\end{table*}
\endgroup

\subsubsection{Correlation measure} \label{uncertainty_estimations}

Our aim is to compute the cross correlation coefficient between the extragalactic foreground models (Figure \ref{fig:models_param}) and the $H_0$ and $\Omega_ch^2$ cosmological parameters maps of FG2021 (two lower panels of Figure \ref{fig:relation_h13_forg}), using the standard definition:
\begin{equation*}
    Corr_{XY} = \frac{\left\langle (X-\overline{X}) \cdot (Y-\overline{Y}) \right \rangle }{\sqrt{\left \langle (X - \overline{X})^2 \right \rangle \left \langle (Y - \overline{Y})^2 \right \rangle}}
\end{equation*}
where $X$ and $Y$ correspond to the extragalactic foreground models and local cosmological parameter variation maps. This cross correlation averages are calculated over the whole sky pixels maps, with pixel and mean values corresponding to both temperature and cosmological parameters. We notice that given the large extension of both foreground model and cosmological parameter maps, the small pixel size adopted (\texttt{NSIDE=4}) is not likely to affect the cross correlation values.

The significance of this cross correlation coefficient will be address by suitable uncertainty estimations as follows. We consider two procedures: firstly, we correlate the extragalactic foreground model with the set of 300 cosmological parameter simulations studied in FG2021 (see their Section 3.7 for details). Secondly, we correlate the actual CMB causal horizons maps with a set of 300 foreground model maps based on mocks of nearby spiral galaxies extracted from the MultiDark Planck 2 (MDPL2) simulation \cite{Multidark} (See Appendix \ref{Mock galaxies} for details) instead of real galaxy data.

\section{\label{sec:Results} Results}

\subsection{Parameter estimation Results} \label{parameter_Correlation_results}

In this section we apply the methodology described and validated in section \ref{sec:metho} to our CMB Planck PR4 masked maps with the final purpose of obtaining a new set of cosmological parameters. This set will be compared with the best-fitting $\Lambda$CDM parameters to assess whether there are significant fluctuations.  Table \ref{tab:parameters} shows the results of our parameter estimation procedure for the set of proposed masks. Together with the basic flat-space $\Lambda$CDM cosmological parameters, we estimate the derived parameter $\sigma_8$ which is a measurement of the mass density fluctuations in a radius of 8 Mpc.

We see that regardless of the mask used, the cosmological parameters do not vary significantly with respect of those obtained by Planck PR3 \cite{PlanckPR3VI} (Table \ref{tab:pipe_valid}) within the larger uncertainties of the estimates on smaller sky fractions. Particularly, considering the tensions in $H_0$, $\Omega_ch^2$ and $\sigma_8$, our results are in agreement with those obtained with the Planck PR3 CMB map. Furthermore, the fact that the foregrounds are significantly more pronounced in dense galactic environments (L2023, GL2023) should imply that the main signal is dominated by the large scales of the nearby galactic filaments which typically extend several degrees on the sky (see for instance the model in H2023). This could result in a foreground signal dominated by multipoles $\ell<32$ which are not used in our parameter estimation pipeline. Due to the larger cosmic variance dominated errors in these low multipoles, we expect the impact of the foreground signal of these multipoles on the cosmological parameters would be minimal.

\begingroup
\setlength{\tabcolsep}{2pt} 
\renewcommand{\arraystretch}{1.5} 
\begin{table} 
    \centering
    \begin{tabular}{c|ccc}
    \hline
        Parameter &  Mask 1 & Mask 2 & Mask 3 \\
        \hline
        $\Omega_c h^2$ &  $0.120 \pm 0.002$ & $0.119 \pm 0.002$ & $0.119 \pm 0.003$ \\
        $\Omega_b h^2$  & $0.0220 \pm 0.0002$ & $0.0222 \pm 0.0002$ & $0.0222 \pm 0.0004$ \\
        $H_0$ &  $67.1 \pm 0.9$ & $68 \pm 1$ & $68 \pm 1$ \\
         $n_s$ & $0.963 \pm 0.006$ & $0.966 \pm 0.006$ & $0.967 \pm 0.008$ \\
        $10^9 A_s$ & $2.09 \pm 0.01$ & $2.08 \pm 0.01$ & $2.08 \pm 0.02$ \\
        $A_{PS}$ & $56 \pm 4$ & $58 \pm 5$ & $57 \pm 6$ \\
        $\sigma_8$ & $ 0.809 \pm 0.008$ & $ 0.805 \pm 0.009$ & $ 0.81 \pm 0.01$ \\
        \hline
    \end{tabular}
    \caption{Flat $\Lambda$CDM cosmological parameters estimation for the set of foreground masks. The uncertainties shown correspond to the parameter estimation for 100 simulations.}
    \label{tab:parameters}
\end{table}
\endgroup

\subsection{Correlation Results} \label{Correlation_results}

\begin{figure*} 
    \centering
    \includegraphics[width=0.9\textwidth]{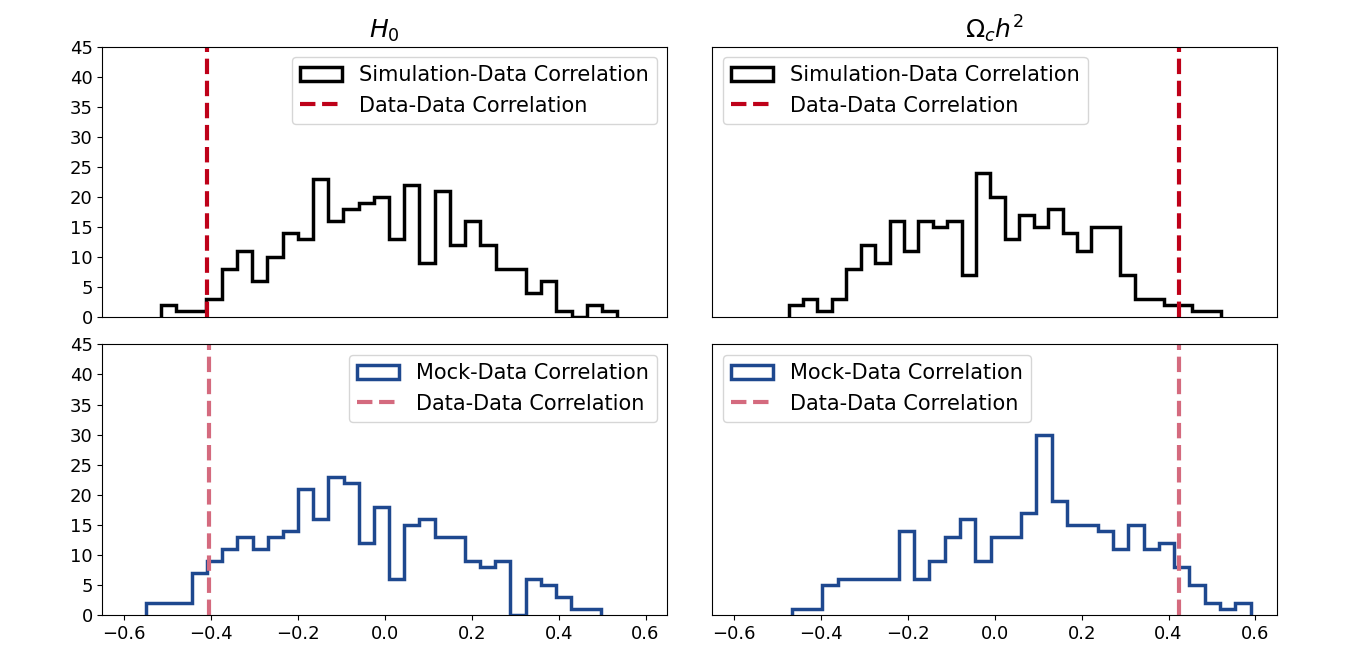}
    \caption{Distributions of the correlation coefficients between extragalactic foreground regions and parameters maps in simulations. Top panels: Parameters simulations method for $H_0$ and $\Omega_ch^2$: the foreground model was correlated with 300 parameter variation maps obtained by local parameter estimates on simulated CMB maps Lower panels: Mock method for both parameters: the Planck CMB map was correlated with 300 maps of the extragalactic foreground based on mocks of a galaxy simulation. In all the panels, the dashed line represents the amplitude of the real data correlation.}
    \label{fig:histograms}
\end{figure*}

Here we will present results for the correlation test detailed in Section \ref{subsec:horizons}. The calculations yield similar results for the three different extragalactic foreground models adopted, giving coefficient values of $[-0.40,-0.50,-0.41]$ and $[0.42,0.53,0.43]$ for $H_0$ and $\Omega_ch^2$ for Model 1, 2 and 3, respectively. Therefore, in the following, uncertainties will only be estimated for Model 3. The results showing the correlation coefficients compared to simulations are shown in Figure \ref{fig:histograms}.

As it can be seen in this figure, the correlations of the two parameters are highly significant at $98.67 \%$ for CMB simulations and $95.67 \%$ for the mock galaxy model. We acknowledge that the more reliable values of uncertainties are those derive from the CMB horizons parameters simulations since the mock galaxy model does not fully mimic the observer properties which is simply centred at random in the galaxy simulation and could induce a large variance in the results. We conclude that we find a correlation between the nearby extragalactic foreground regions and the CMB causal parameters horizons at $2.5\sigma \sim 3 \sigma$ calibrated on 300 simulated maps.

\begin{figure}
    \centering
    \includegraphics[width=\columnwidth]{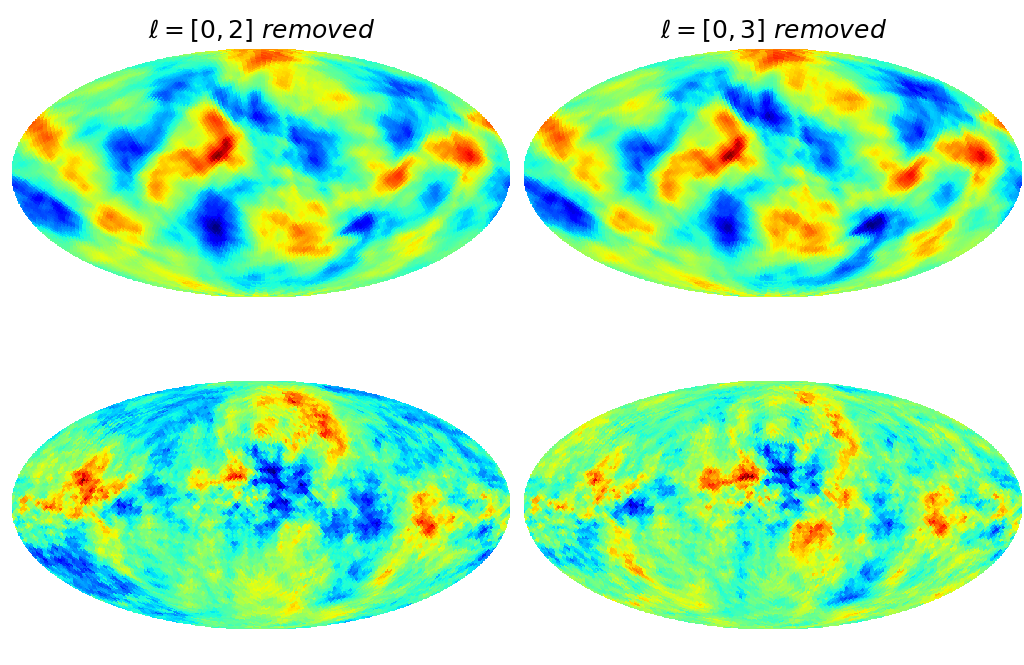}
    \caption{Parameter variation maps and foreground models with quadrupole (left panel) and also octopole (right panel) removed. The parameter variation map is the $\Omega_ch^2$ map from Figure \ref{fig:relation_h13_forg} and the model map is model 3 from Figure \ref{fig:models_param}.}
    \label{fig:removing_multipoles}
\end{figure}

Looking at the parameter variation maps in Figure \ref{fig:relation_h13_forg} and the galactic model maps in Figure \ref{fig:models_param}, they appear to be dominated by a quadrupole moment. In order to test possible smaller scales structures in the parameter variation and foreground model, we remove the quadrupole of both maps. In Figure \ref{fig:removing_multipoles}, we see indeed that smaller scales are revealed when the dominating quadrupole is removed. We also see that further removing the octopole, little more structure is revealed which is not surprising given the large smoothing angle.

\begin{figure*}
    \centering
    \includegraphics[width=\textwidth]{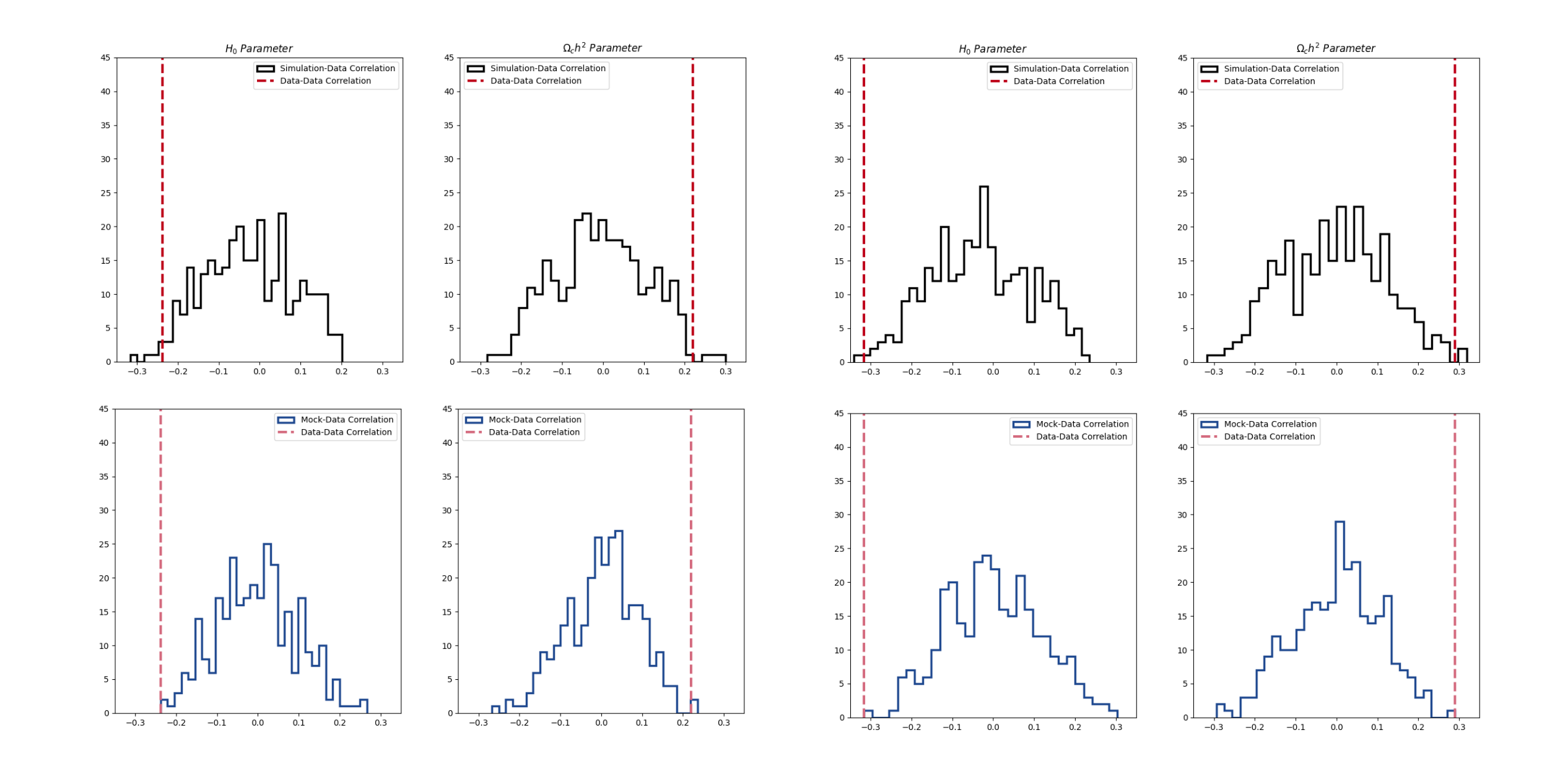}
    \caption{Same as Figure \ref{fig:histograms}, but with the quadrupole and octopole removed before calculating the correlation coefficients. On the left side, the results of removing the quadrupole for both simulations (top panels) and mocks (bottom panels). On the right side, removing also the octopole.}
    \label{fig:removing_l2l3}
\end{figure*}

The results are presented in Figure \ref{fig:removing_l2l3}, where the higher significance for these cases can be seen. The correlation is now even stronger, exceeding $3\sigma$ for both simulations and mocks, assuming a Gaussian distribution and relying on the value of $\sigma$ found from 300 simulations. The quadrupole dominated map gives a large spread in correlation values with simulations. The few degrees of freedom of a quadrupole increases the probability of a chance alignment between the two maps. The octopole with more degrees of freedom reduces this probability and the fact that the significance of the correlation increases, is a strong indication of the new foreground being a probable explanation for the large local variation of the cosmological parameters discovered in FG2021.

\section{\label{sec:Conclusions} Conclusions}

In this work, we have re-estimated cosmological parameters in Planck CMB maps, masking expected areas of high contamination from the new extragalactic foreground around spiral galaxies and also tested whether the unexpectedly high variation of cosmological parameter estimates across the sky of FG2021 could be related to this new foreground.

After validating the results of out pipeline with official Planck cosmological parameter estimates, we find that different attempts of masking the new foregrounds in general gives parameter estimates consistent with the full-sky results. We acknowledge however that the use of a reduced sky area could make possible differences to be within the larger uncertainties obtained.

Our second analysis on the other hand, considers a modelling of the extragalactic foreground which can be used to study the relation with the causal horizons studied in FG2021. This allows for the use of the full sky (outside the common mask), instead of roughly $50 \%$ of the sky used when the extragalactic foreground masks are applied. As shown in our work, even with a very simplified model, which only partially reproduces the observed foreground and only takes into account galaxies with $z<0.02$, we find a significant correlation with the local parameter variation maps at the $2-3\sigma$ confidence level. Nevertheless, we see that this correlation accounts for $(40-50)\%$ correlation with $15\%$ uncertainty of the effect, and it is not sufficient to fully explain the existence of the horizons (see references below \cite{Gaztañaga1,Gaztañaga2,Gaztañaga3,Gaztañaga4,Gaztañaga5}). We argue that the areas on the sky with extreme parameter values which can be interpreted as casual horizons are largely aligned with the presence of large spiral galaxies and could be explained by the new foreground. This remarkable coincidence between anomalous CMB regions and nearby galaxies provides an independent test of the new extragalactic foreground, as well as its possible influence on the inference of cosmological parameter values.

More work is needed to explore the impact of these new foregrounds on other very large-scale CMB anomalies, such as the parity asymmetry at $\ell <20$, which hint at new physics on super-horizon scales and can be used to explain the anomaly of the low-quadrupole and hemispherical power asymmetry \cite{2024JCAP...06..001G}.

\begin{acknowledgments}
This work was partially supported by Agencia Nacional de Promoción Científica y Tecnológica (PICT 2015-3098, PICT 2016-1975), the Consejo Nacional de Investigaciones Científicas y Técnicas (CONICET, Argentina) and the Secretaría de Ciencia y Tecnología de la Universidad Nacional de Córdoba (SeCyT-UNC, Argentina). Results in this paper are based on observations obtained with Planck (\url{http://www.esa.int/Planck}), an ESA science mission with instruments and contributions directly funded by ESA Member States, NASA, and Canada. The simulations were performed on resources provided by UNINETT Sigma2 - the National Infrastructure for High Performance Computing and Data Storage in Norway". Some of the results in this paper have been derived using the HEALPix package. \citep{healpix}
This work used computational resources from CCAD – Universidad Nacional de Córdoba (\url{https://ccad.unc.edu.ar/}), which are part of SNCAD – MinCyT, República Argentina
\end{acknowledgments}

\appendix
\section{CMB simulations and Mocks} \label{Mock galaxies}
\begin{figure} [H]
    \centering
    \includegraphics[width=\columnwidth]{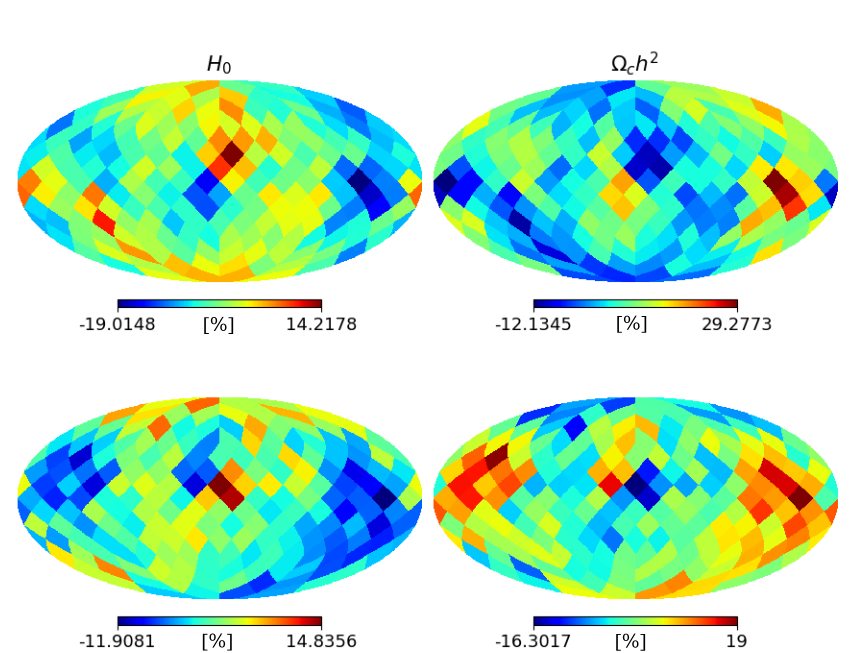}
    \caption{Typical simulated CMB cosmological parameters variation maps from FG2021. For simulated CMB maps, the cosmological parameters were estimated locally on $60^\circ$ discs in the same way as for the Planck CMB map. Here we see the resulting parameter maps. The left panels show two $H_0$ variation maps while the right panels are maps of the local variations of the $\Omega_ch^2$ parameter. The resolution is \texttt{NSIDE=4}.}
    \label{fig:simulations_comparation}
\end{figure}

For the estimation of uncertainties in the correlation coefficients between our galaxy based foreground model and the local parameter variation maps, we have used two different methods, as described in Section \ref{uncertainty_estimations}. The first procedure consists in the correlation of our extragalactic foreground Model 3 with a suite of 300 simulated parameter variation maps from FG2021, constructed in the same way as the parameter map for the actual Planck CMB map, but based on Gaussian simulations: As detailed in FG2021, these maps include the effect of CMB lensing, a residual foreground sky and an added Poisson-like ‘residual foreground’ amplitude that effectively accounts for the combination of Cosmic Infrared Background (CIB) and extragalactic point sources contributions from different frequency channels. We restrict the analysis to the simulated parameter maps for the $H_0$ and $\Omega_ch^2$ parameters. The maps corresponding to two typical simulations are shown in Figure \ref{fig:simulations_comparation}.

\begin{figure} [H]
    \centering
    \includegraphics[width=\columnwidth]{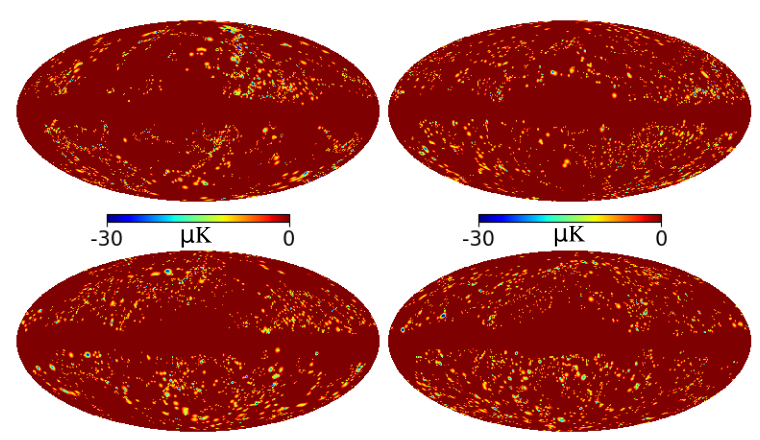}
    \caption{Three mock foreground models and the actual foreground model from observations (Top left panel). Although the actual temperature amplitude of the models are highly uncertain, we only use these models for cross-correlations with parameter variation maps where the values are normalized to their standard deviations. }
    \label{fig:mocks_comparation}
\end{figure}

The second procedure consists in the calculation of the correlation between the actual parameter variation map and  a set of 300 foreground models based on mocks of nearby galaxies. With the aim to develop a suitable set of simulated nearby galaxies for the analysis, we make use of the MDPL2 simulation  \cite{Multidark}. In order to reproduce reliably a galaxy distribution with similar properties as our local Universe and the actual extragalactic foreground, for each observer we restrict the maximum distance to the galaxy ($D \leq 80$Mpc/h) and we use the $g-r$ colour to separate elliptical and spiral galaxies. Since the extragalactic foreground is associated to spiral galaxies, we use them to generate the mock foreground maps through Model 3 as explained in Section \ref{foreground_models}. In Figure \ref{fig:mocks_comparation} we show the foreground models derived from three typical galaxy mocks.

\bibliography{biblio}

\end{document}